\documentclass[a4paper]{article}

\usepackage{Odyssey2018}

\usepackage{amsmath,graphicx}
\usepackage[table]{xcolor}
\usepackage{booktabs}
\usepackage{amssymb,bm}
\usepackage{textcomp}
\usepackage{multirow}

\ninept
\newcommand{\EE}       {\boldsymbol{\Sigma}}
\newcommand{\mm}       {\boldsymbol{\mu}}

\title{Spoken Pass-Phrase Verification in the i-vector Space}

\makeatletter
\def\name#1{\gdef\@name{#1\\}}
\makeatother
\name{{\em Hossein Zeinali\,$^{1,2,3}$, Luk\'{a}\v{s} Burget\,$^2$, Hossein Sameti\,$^1$, Jan ``Honza'' \v{C}ernock\'{y}\,$^2$}}

\address{
  $^1$ Sharif University of Technology, Tehran, Iran \\
  $^2$ Brno University of Technology, Speech@FIT and IT4I Center of Excellence, Czech Republic \\
  $^3$ Sharif DeepMine Ltd., Tehran, Iran \\
  {\small \tt hsn.zeinali@gmail.com, burget@fit.vutbr.cz, sameti@sharif.edu, cernocky@fit.vutbr.cz}
}

\begin{document}

\maketitle

\begin{abstract}
The task of spoken pass-phrase verification is to decide whether a test utterance contains the same phrase as given enrollment utterances. Beside other applications, pass-phrase verification can complement an independent speaker verification subsystem in text-dependent speaker verification. It can also be used for liveness detection by verifying that the user is able to correctly respond to a randomly prompted phrase. In this paper, we build on our previous work on i-vector based text-dependent speaker verification, where we have shown that i-vectors extracted using phrase specific Hidden Markov Models (HMMs) or using Deep Neural Network (DNN) based bottle-neck (BN) features help to reject utterances with wrong pass-phrases. We apply the same i-vector extraction techniques to the stand-alone task of speaker-independent spoken pass-phrase classification and verification. The experiments on RSR2015 and RedDots databases show that very simple scoring techniques (e.g. cosine distance scoring) applied to such i-vectors can provide results superior to those previously published on the same data.
\end{abstract}

\section{Introduction}
\label{sec.intro}

Utterance Verification (UV) is the task of confirming the content of a given utterance and answering the question of whether the user uttered the prompted pass-phrase or not. In this paper, we focus on spoken pass-phrase verification, where one or more spoken examples are given for the required pass-phrase. In other words, the task is to verify whether a test utterance (from, possibly, a previously unseen speaker) contains the same pass-phrase as a given enrollment utterance or a set of enrollment utterances. UV is a subtask of text-dependent Speaker Verification (SV), where the correctness of the uttered pass-phrase needs to be verified together with the speaker identity. Here, the UV component helps to prevent from replay attacks using a random utterance of the target speaker.

A single model is often used in text-dependent SV to jointly address both UV and SV tasks. For example, in our previous work on i-vector based text-dependent SV~\cite{zeinali2016trans, zeinali2016ivector, zeinali2015telephony}, HMMs constructed specifically for each pass-phrase were used to extract sufficient statistics in order to make the resulting i-vectors both speaker and phrase specific. Further text-dependent SV experiments have shown that it is enough to use the more conventional Universal Background Model - Gaussian Mixture Model (UBM-GMM) if the i-vectors are extracted from BN features~\cite{zeinali2016deep, zeinali2016csl}. The frame-by-frame BN features are obtained from a DNN, which is trained to extract phonetic information from the acoustic context (300 ms) around the current frame. I-vectors extracted from such BN features contain lots of information about the phonetic content of the corresponding utterances and are very good for rejecting utterances with incorrect pass-phrases.

Although a single model can be used to jointly address both UV and SV tasks, there is still good reason to have stand-alone (speaker-independent) system for utterance verification or classification. In text-dependent SV, for example, replay attacks with pre-recorded correct pass-phrases are very difficult to reject. A possible way to tackle this problem is  to use anti-spoofing techniques based on detecting typical distortions in recorded and replayed audio~\cite{kinnunen2017asvspoof} or using audio fingerprinting~\cite{wu2014study} to detect a replay of an enrollment utterance. However, these techniques are often not very reliable. An alternative is to use a liveness detection using a separate UV subsystem as follows: in one step, a standard text-dependent SV is used to verify the speaker, while in the second step the user is prompted some random phrase, which he needs to pronounce to prove his responsiveness. Speaker identity can be verified from this second phrase in more text independent fashion. More importantly, the correctness of the phrase can be verified by the UV subsystem. The prompted random phrase can be in a textual form or can be represented by audio recording. The later case is of our main interest. Note that the UV techniques can be also applied to other problems than the text-dependent SV. An example can be re-scoring detections in keyword spotting or query-by-example system~\cite{Dai2005}.

In this work, we experiment with the aforementioned i-vector based text-dependent SV techniques. However, we apply these techniques to the stand-alone task of (speaker-independent) spoken pass-phrase verification or classification. We show that the i-vectors extracted in the described way contain predominantly information about the lexical content of the utterance and are therefore excellent representations for this task. We also show that our solution based on the simple i-vector representation outperforms the previously proposed and often more computationally complex methods, which serve as our baseline~\cite{kinnunen2016utterance}.

\section{Baseline Utterance Verification Methods}
\label{sec.utter_verif}

The effort on UV described in the literature is quite limited. In~\cite{kinnunen2016utterance}, four systems are described, which constitute a good example of the standard techniques for UV. We use these systems and the corresponding results as our baseline. In some of our experiments with i-vector based UV, the same setup as in~\cite{kinnunen2016utterance} is used to make the results directly comparable. Here, we provide the only brief description of these baseline systems. For more detailed description, we kindly refer the reader to~\cite{kinnunen2016utterance}.

The system denoted as {\bf UV1} uses Mel-Frequency Cepstral Coefficients (MFCCs) with their first and second order derivatives and a GMM-UBM with 512 Gaussian components trained on TIMIT data. The utterance models are adapted from the GMM-UBM using the standard relevance maximum-a-posteriori (MAP) adaptation~\cite{reynolds2000speaker} and the log-likelihood ratio between the utterance and the UBM serves as the UV score. Note that this technique only models the distribution of acoustic features in the training utterances, but does not try to model the temporal structure of the uttered phrases.

The system denoted as {\bf UV2} uses 5-state HMM with the left-to-right topology to model the temporal structure of utterances. Each state is modeled using a GMM, which is MAP adapted in a similar manner and from the same GMM-UBM as in the case of the system UV1. Viterbi alignment of frames to HMM states is used to train phrase specific models on training utterances and to evaluate the log-likelihood ratio score for the test utterances.

The {\bf UV3} system uses perhaps the most conventional approach to spoken utterance verification: dynamic time warping (DTW)~\cite{rabiner1993fundamentals} is used to frame-align utterances and to calculate the distances between the utterances. Euclidean distance between MFCC feature vectors is used as the frame-to-frame distortion. Note that the DTW based UV could be further improved by using more sophisticated frame-to-frame distortions~\cite{boulianne2015language} or by calibrating the resulting DTW scores to make them proper UV log-likelihood ratios~\cite{liu2010utterance}. These improvements are, however, not considered in this work.

{\bf UV4} makes use of a DNN based automatic speech recognition (ASR) system trained on TIMIT data using Kaldi~\cite{kaldi2011} toolkit. Each test utterance is forced-aligned to the known reference transcript of a given pass-phrase and the acoustic score (pseudo log likelihood) for this alignment is used as the UV score. Note that this system performs UV using the pass-phrase given as text, unlike the other methods described in this paper, which rely on spoken pass-phrase.

\section{i-vector Based Utterance Verification}
\label{sec.methods}

In this work, we use i-vectors as fixed length low-dimensional representations of speech utterances. First, i-vectors were proposed for the task of text-independent speaker recognition~\cite{dehak2011front}, but soon became popular for other tasks of utterance level classification or verification such as language, gender, signature, age or emotion recognition~\cite{martinez2011language, zeinali2017usage, bahari2012age, xia2012using}. In the probabilistic model for i-vector extraction, a low-dimensional latent variable is used to representing utterance specific GMM. I-vector is the MAP point estimated of the latent variable adapting the corresponding GMM to a given speech utterance. For more details on the i-vector model, we kindly refer the reader to other sources~\cite{dehak2011front, zeinali2016csl}. Here, we only recall that the i-vector can be inferred from sufficient statistics, which are collected from the speech utterance. To collect the sufficient statistics, we need an alignment of speech frames to i-vector model Gaussian components. This alignment is traditionally obtained using an underlying UBM-GMM.

\subsection{HMM based frame alignment methods}

In our previous works on text-dependent SV~\cite{zeinali2016trans, zeinali2016ivector} and also text-prompted SV~\cite{zeinali2015telephony}, i-vectors were extracted using HMM based alignment. For this purpose, phoneme recognizer is first trained, where mono-phone 3-state HMMs are used with state distributions modeled using GMMs. Given the known transcriptions of enrollment and test utterances, the phrase specific HMMs are constructed from the mono-phone HMMs. The Viterbi algorithm is then used to obtain the alignment of the frames to the HMM states in order to collect the sufficient statistics. Note that, while there is a specific HMM built for each phrase, there is only one set of Gaussian components (Gaussians from all the HMM states of all phone models) corresponding to a single phrase-independent i-vector extraction model. The i-vector extractor is trained and used in the usual way, except that, it benefits from the better alignment of frames to Gaussian components as constrained by the HMM model. More details on this i-vector extraction method can be found in~\cite{zeinali2016trans, zeinali2016csl}.

For text-dependent SV, it was shown~\cite{zeinali2016trans, zeinali2016ivector} that this alignment extraction strategy produces more phrase specific i-vectors, which are especially effective for rejecting utterance with wrong pass-phrases. For the same reason, this technique is also suitable for utterance verification task as demonstrated in our experiments. One the drawback of this approach is that we need to know the phrase specific phone sequence for constructing the corresponding HMM.

\subsection{Bottleneck features}

MFCCs were conventionally used as the speech features for i-vector extraction. More recently, however, significant improvements were obtained for both text-dependent~\cite{zeinali2016deep, zeinali2016csl} and text-independent~\cite{lozano2016analysis, tian2015investigation, matejka2016analysis} verification task from using BN features or concatenated MFCC+BN features. Note that BN features were previously successfully used also in other areas of speech processing~\cite{grezl2009investigation, yaman2012bottleneck, vesely2012language}.

BN features are frame-by-frame extracted using a bottleneck DNN, which is typically trained for phone classification. Bottleneck DNN is a neural network with a specific topology, where one of the hidden layers has significantly lower dimensionality than the surrounding layers. A bottleneck feature vector is generally understood as a by-product of forwarding a primary input feature vector through the DNN, while reading the output of the bottleneck layer where the relevant information is compressed into a low dimensional vector. In this work, we use more elaborate architecture for BN features called Stacked Bottleneck Features~\cite{karafiat2014but}. This architecture is based on a cascade of two such BN DNNs. The BN output of the first network is \emph{stacked} in time, defining context-dependent input features for the second DNN. The input features to the first stage DNN are 36 log Mel-scale filter bank outputs augmented with 3 fundamental frequency features~\cite{karafiat2014but} and normalized using conversation-side based mean subtraction. The outputs from the BN layer of the second stage DNN are then taken as the final output features (i.e. the features to train the i-vector model on). With this architecture, each output feature vector is effectively extracted from at least 30 frames (300 ms) of the input features in the context around the current frame. Therefore, each BN feature vector contains important information about the phonetic context around the current frame, which is further propagated to the i-vector extracted from these features. This makes BN feature based i-vectors very phrase specific even when extracted using the conventional UBM-GMM model (i.e. there is no need for the HMM based alignment), which was previously demonstrated in text-dependent SV experiments~\cite{zeinali2016deep, zeinali2016csl}.

\subsection{Scoring methods}

In our experiments, we consider both the task of close-set pass-phrase classification and open-set pass-phrase verification. To classify or compare i-vectors, we use only two very simple techniques, namely Linear Gaussian Classifier (LGC) and cosine similarity scoring.

\subsubsection{Linear Gaussian Classifier (LGC)}
\label{ssec.lgc_method}

For each class (pass-phrase) $i = 1 \dots K$, LGC assumes Gaussian distribution of i-vectors $\mathcal{N}(\mathbf{w} | \mm_{i}, \EE)$. Each class is modeled by its own mean vector $\mm_{i}$. All the classes, however, share the same average within-class covariance matrix $\EE$, which is typically estimated as
\begin{eqnarray}
    \mm_i & = & \frac{1}{N_i}\sum_{n=1}^{N_i}\mathbf{w}_i^n    \\
    \label{eq.within_cov}
    \EE & = & \frac{1}{N}\sum_{i=1}^{K}\sum_{n=1}^{N_i}
    (\mathbf{w}_i^n - \mm_i)(\mathbf{w}_i^n - \mm_i)^T \,,
\end{eqnarray}
where $N_i$ is the number of training samples (i-vectors) for phrase $i$ and $\mathbf{w}_i^n$ is the $n^{\mathrm{th}}$ training sample of phrase $i$. Once the model is trained on the training (or enrollment) utterances, evaluation data can be classified by simply selecting the class with the highest posterior probability:
\begin{equation}
	\label{eq.likelihood_ratio_lgc}
	P\:(i|\mathbf{w}) = \frac{\mathcal{N}(\mathbf{w}|\mm_{i}, \EE)P(i)}{\sum_{k=1}^{K} \mathcal{N}(\mathbf{w}|\mm_{k}, \EE)P(k)}\:,
\end{equation}
were we assume equal priors $P(i)$ for all classes. To be consistent with results from~\cite{kinnunen2016utterance}, we also report the performance in terms of Equal Error Rate (EER) for LGC, where the posterior probabilities serve as the verification score for the corresponding classes. In this case, however, we cannot talk about open-set verification as the score from the close-set of $K$ phrases depends on each other through the normalization in the posterior probability calculation.

\subsubsection{Cosine Similarity Scoring}
\label{ssec.cosine_method}

Cosine similarity scores are also used in our experiments to perform classification and verification of i-vectors. In this case, the enrolled pass-phrase models are obtained as a simple average of training (or enrollment) i-vectors. Note that there is no need to estimate any covariance matrix for this scoring method, which makes it more robust for the cases where only few training examples are available. To perform classification of a test utterance, we can select the class with the highest cosine similarity score. For the detection task (i.e. to evaluated EER), we simply use the cosine similarity score as the verification scores. Note that in this case, verification scores for individual pass-phrases are completely independent of each other and the obtained EER can be correctly interpreted as open-set pass-phrase verification performance.

Again, to be consistent with results from~\cite{kinnunen2016utterance}, we alternatively normalize the cosine similarity scores using the so-called Max-Norm method. In this case, for each test utterance, the maximum of cosine scores over all other the $K-1$ phrases is subtracted from the original cosine scores. The same normalization is also used for some of the results from~\cite{kinnunen2016utterance}, which are also presented for comparison in Table~\ref{tbl.reddots_results}. Although the normalization (seemingly) improves the classification and verification results, we no more deal with the open-set verification problem just like in the case of LGC.

\subsubsection{Motivation for simple classifiers}

\begin{figure}[t]
	{\centering \includegraphics[width=8.0cm,trim={10mm 10mm 10mm 10mm}, clip]{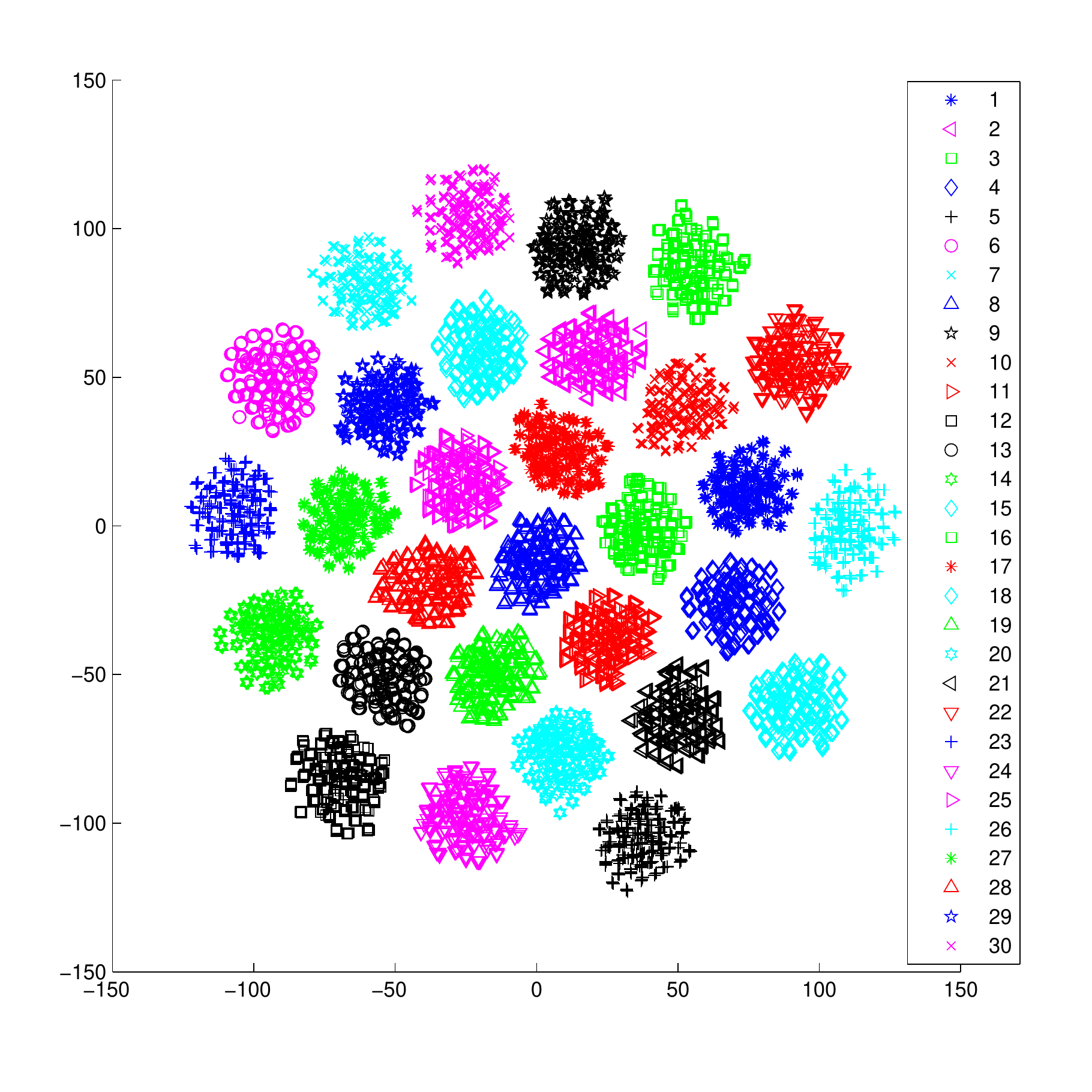}}
	\vspace{-4mm}
	\caption{Male i-vectors from the RSR2015 database reduced to 2-dimensional space using t-SNE. There are 30 well separated clusters corresponding to different phrases.}
	\label{fig.tsne_male}
\end{figure}

We have used t-SNE~\cite{maaten2008visualizing} to reduce 400-dimensional i-vectors extracted using UBM-GMM from MFCC+BN features into 2-dimensional space. The i-vectors were taken from all male speakers from the RSR2015 test set. \figurename~\ref{fig.tsne_male} shows the plot of the resulting vectors for 30 phrases of the RSR2015 database. Each point in the plot corresponds to one i-vector and is colored according to the phrase label. One can see that i-vectors from different phrases form nicely separated clusters in the t-SNE space. Moreover, all classes have roughly Gaussian distribution with the same within-class covariance matrix. Although, in general, t-SNE provides nonlinear transformation of the original space, the nicely separated clusters and simple distributions make us believe that pass-phrase verification should be an easy task in this i-vector space and simple scoring technique should be sufficient.

\section{Experimental setup}
\label{sec.setup}

We report results on Part1 of the RSR2015 database~\cite{larcher2014text} as well as Part1 of the RedDots database~\cite{lee2015reddots}. RSR2015 comprises recordings from 30 different phrases. The same phrases appear in three disjoint subsets of speakers {\em background}, {\em development} and {\em evaluation set}. Each speakers repeats each phrase 9 times. The male utterances from the {\em background set} (50 speakers) are used for training the classifiers. The results are reported on male part of the {\em evaluation set} (57 speakers). The {\em development set} is not used in our experiments.

Part1 of RedDots contains 49 male speakers, each pronouncing several times 10 common pass-phrases. For the results reported in Table~\ref{tbl.reddots_results}, the evaluation setup defined at UEF university was adopted to make our results directly comparable with those previously reported in~\cite{kinnunen2016utterance}. In this setup, all utterances from 9 different speakers are used to train the 10 pass-phrase models (In total, 1485 utterances are used for training, roughly 148 utterances per pass-phrase) and 30 other speakers are selected for the evaluation set. Note that 10 out of the 49 male speaker were not used at all in our experiments. For the results reported in Table~\ref{tbl.eers} analyzing the performance for reduced amount of training data, several subsets of the 9 training speakers are used and evaluation set remains unchanged.

A UBM-GMM with 1024 components, alignment HMMs with 3 states and 8 components in each state  and a 600-dimensional i-vector extractor are trained using LibriSpeech database~\cite{panayotov2015librispeech} and the {\em background set} of RSR database. BN feature extractor was also trained on LibriSpeech~\cite{zeinali2016csl}.

In order to evaluate EERs, the verification scores for the individual pass-phrases are simply polled. We understand that it is questionable to use such pooled EER in the case of close-set problem, where the verification scores are normalized using the scores from the competing hypothesis (i.e. our results with LGC and Cosine Max-Norm). Nevertheless, we also include such results in order to allow for comparison with the baselines from~\cite{kinnunen2016utterance}. For the RSR2015 database, each evaluation utterance forms one target trial and 29 non-target trials corresponding to the remaining pass-phrases. Similarly, one target and 9 non-target trials are formed for each evaluation utterance from RedDots.

%

\section{Results}
\label{sec:results}

\subsection{RSR2015 Results}

First, we report results on male utterances from RSR2015 {\em evaluation set}. This is an example of a scenario where plenty of training examples are available for each of 30 pass-phrases ($50 \times 9=450$ training utterances per phrase). Further, we deal here with the ideal condition, where UBM and i-vector extractor are trained on training data of the same phrases. This leads to nearly faultless recognition performance for any of the scoring technique as presented in Table~\ref{tbl.methods_norm_comp}. In this case, we have chosen i-vectors extracted using UBM-GMM from MFCC+BN features, which was the configuration previously providing excellent performance in text-dependent SV task~\cite{zeinali2016deep, zeinali2016csl}.

\begin{table}[t]
	\renewcommand{\arraystretch}{1.0}
	\caption{\label{tbl.methods_norm_comp} Performance of the i-vector based methods on RSR2015 .}
	\vspace{3mm}
	\centering{
		\setlength\tabcolsep{5pt}
		\begin{tabular}{ l c c}
			\toprule
            \midrule
						        & Classification	& EER\\
						        & Error [\%]	    & [\%]\\
            \midrule
			LGC			        & 0.0 		& 0.000 \\
			Cosine		        & 0.0 		& 0.007 \\
			Cosine Max-Norm		& 0.0 		& 0.000 \\
			\midrule
		\bottomrule
		\end{tabular}
	}
\end{table}

\subsection{RedDots Results}

Table~\ref{tbl.reddots_results} shows results for more challenging RedDots database. The phrase models are still trained on relatively many examples. As mentioned in section ~\ref{sec.setup} there are about 148 examples from 9 different speakers for each pass-phrase. But the data are recorded under more challenging conditions and the UBM and i-vector extractor are trained on data of mismatched phrases. Note also that, in the case of LGC, within-class covariance matrix (i.e. Eq. \eqref{eq.within_cov}) was estimated on RSR2015 on data of different phrases (i.e. in a phrase independent fashion). Only the class means were estimated on RedDots data.

The first section of Table~\ref{tbl.reddots_results} shows results obtained with the baseline systems, which were described in Section~\ref{sec.utter_verif}.
These results are borrowed from Table 5 of~\cite{kinnunen2016utterance} and are directly comparable with our result from the second section of Table~\ref{tbl.reddots_results}.
The results show that the proposed i-vectors (again UBM-GMM and BN features are used) easily outperform even the fusion of the previously published baseline methods from~\cite{kinnunen2016utterance}. We have manually inspected the utterances where the i-vector based systems made an error, and we have observed that those were most severely corrupted utterances (i.e. mispronunciation, only silence, etc). Note also the very good performance of the Cosine similarity with no normalization, which is the result for the true open-set pass-phrase verification task.

\begin{table}[t]
	\renewcommand{\arraystretch}{1.0}
	\caption{\label{tbl.reddots_results} Comparison of the i-vector bases methods with the baseline methods from~\cite{kinnunen2016utterance} on RedDots data.}
	\vspace{3mm}
	\centering{
		\setlength\tabcolsep{3pt}
		\begin{tabular}{ l c c c c }
			\toprule
			\midrule
			Method			& & No-Norm & Max-Norm & Classification \\
			    			& & EER [\%] & EER [\%] & Error [\%] \\
			\midrule
			UV1							& & 9.31 & 2.08 & -- \\
			UV2							& & 5.54 & 1.11 & -- \\
			UV3							& & 24.81 & 7.80 & -- \\
			UV4							& & 16.60 & 4.56 & -- \\
			Fused\,(UV1 \dots UV4)		& & 6.13 & 1.43 & -- \\
			\midrule
			LGC							& & 0.11 & --   & 0.25 \\
			Cosine						& & 0.61 & 0.10 & 0.25 \\
			\midrule
			\bottomrule
		\end{tabular}
	}
\end{table}

From the results in Table~\ref{tbl.reddots_results}, we can see that both LGC and Cosine distance perform similarly. This is understandable realizing the close relation between the two scoring methods: LGC with identity within-class covariance matrix applied to length normalized i-vectors\footnote{However, note that we do not apply the length normalization in the case of LGC scoring in our experiments.} would produce class likelihood  proportional to Cosine distance. In reality, the within-class covariance matrix will not be far from identity as the i-vector extractor is trained to produce standard normal distributed i-vectors. Moreover, the Max-Norm applied to Cosine distance scores can be seen as an approximation to the softmax normalization embedded in equation~\ref{eq.likelihood_ratio_lgc}.

\subsection{Features, Alignments and Amount of Training Data}

\begin{table}[t]
	\renewcommand{\arraystretch}{1.0}
	\caption{\label{tbl.eers} Comparison of features, alignment methods and different amount of training examples on RedDots. Three training i-vectors are used per speaker. The results are EERs [\%]}
	\vspace{3mm}
	\centering{
		\setlength\tabcolsep{2pt}
		\begin{tabular}{ l c c c c c c c }
			\toprule
			\midrule
				& & & \multicolumn{5}{c}{Number of Speakers} \\
				\cmidrule{3-8}
				Method 	& Feature\,/\,Align & & \bf1 & \bf2 & \bf3 & \bf5 & \bf9 \\
				\midrule
                		& MFCC\,/\,GMM    & & 61.01 & 7.78 & 3.70 & 2.71 & 1.45 \\
               	LGC		& MFCC\,/\,HMM    & & 9.60  & 1.55 & 1.16 & 1.15 & 0.85 \\
                		& MFCC+BN\,/\,GMM & & 39.11 & 1.10 & 0.21 & 0.15 & 0.14 \\
               	\midrule
                		& MFCC\,/\,GMM    & & 24.54 & 16.7 & 12.9 & 10.1 & 7.17 \\
               	Cosine  & MFCC\,/\,HMM    & & 19.19 & 9.58 & 7.18 & 4.87 & 3.02 \\
                		& MFCC+BN\,/\,GMM & & 7.53  & 2.00 & 1.35 & 0.95 & 0.55 \\
               	\midrule
               	\multirow{3}{*}{\begin{tabular}{@{}l@{}}Cosine \\ Max-Norm\end{tabular}}
                		& MFCC\,/\,GMM    & & 15.51 & 8.18 & 5.67 & 3.62 & 2.01 \\
                		& MFCC\,/\,HMM    & & 9.79  & 3.51 & 2.36 & 1.16 & 0.50 \\
                		& MFCC+BN\,/\,GMM & & 2.52  & 0.35 & 0.30 & 0.20 & 0.10 \\
			\midrule
			\bottomrule
		\end{tabular}
	}
\end{table}

Table~\ref{tbl.eers} compares results obtained with the different proposed i-vector extraction variants: UBM-GMM vs. HMM alignment, MFCC vs. MFCC+BN features. For LGC, within-class covariance matrix was again estimated on RSR20105 data. The results show the degradation of the performance with the decreasing number of training examples (and training speakers). Here, we use only three training examples per speakers and the columns of the table correspond to the number of speakers considered for the training. With only MFCC features, HMM alignment performs better than UBM-GMM in almost all cases. This is due to the HMM ability to model the temporal structure of individual phrases, which has been previously shown to be very effective for rejecting the wrong phrase trials~\cite{zeinali2016deep, zeinali2016trans}. The best performance is achieved with MFCC+BN with UBM-GMM alignment. In this case, the information about the temporal structure of phrases is encoded directly in the BN features, which are extracted from a considerably large context window (i.e. more than 300\,ms). This allows us to obtain the superior performance even with the simpler UBM-GMM alignment.

Again, excellent results can be obtained with MFCC+BN features and the simple Cosine similarity scoring without any normalization, considering that this corresponds to open-set verification task. In this case, acceptable performance is achieved just with 3 samples from 5 speakers (still outperforming all the baseline systems from Table~\ref{tbl.reddots_results}), which might lead to very useful and practical applications.

The simplicity of the i-vector based scoring methods and the relatively low number of parameters that need to be estimated on the training data of matching pass-phrases makes our approach suitable also for the cases with very limited amount of training examples. As can be seen from the results, acceptable performance can be obtained even with only 2 training speakers.

In the case of only single enrollment speaker, the results for LGC based scoring seems to be quite unstable as compared to Cosine distance (e.g. note the surprisingly high 39.11\% EER for MFCC+BN\,/\,GMM). Our further analysis revealed that this was due to the insufficient data used for the estimation of the LGC within-class covariance matrix. As mentioned above, the covariance matrix is pre-estimated on the RSR2015 data of mismatched pass-phrases. Estimating the covariance matrix on more (still mismatched) data helped alleviated this problem.

\section{Conclusions}
\label{sec.conc}

In this paper, we proposed simple but effective i-vector based spoken pass-phrase verification methods and evaluated them on two standard databases: RSR2015 and RedDots. Experimental results have shown the effectiveness of the methods, which achieved almost zero error rate on both databases and significantly outperformed previously published result. 

The main reason for the excellent performance of these methods is the suitability of i-vectors for utterance verification. I-vector extracted from short duration utterance contains predominantly information about the phonetic content of the utterance. Therefore, such i-vectors naturally form phrase specific cluster in the i-vector space without any need for channel compensation and score normalization~\cite{zeinali2016trans, zeinali2016deep}, which are otherwise necessary for tasks like speaker verification.

The advantages of the proposed methods are simplicity, speed, very low overhead and excellent performance. Another interesting property is suitability of these methods for low resource scenarios, which is allowed by their good performance with little amount of training data.

Although the proposed methods have achieved near zero error rate on both databases, we can hardly say that the pass-phrase verification is a solved problem. Much larger databases with plenty of phrases will be necessary to reliably evaluate the verification methods and also to analyze the possible performance degradations due to the phrase similarity. This is an open topic for future works.

\section{Acknowledgment}

The work was supported by Czech Ministry of Education, Youth and Sports from Project No. CZ.02.2.69/0.0/0.0/16\_027/0008371 and the National Programme of Sustainability (NPU II) project "IT4Innovations excellence in science - LQ1602" and also partially supported by Sharif DeepMine Ltd. company in Iran.


\bibliographystyle{IEEEbib}
\bibliography{Speaker,UtterVerif}

\end{document}